\documentclass[runningheads]{llncs}

\usepackage{cite}
\usepackage{amssymb,amsfonts}
\usepackage{amsmath,amssymb}
\usepackage{algorithm}
\usepackage{algorithmic}
\usepackage{graphicx}
\usepackage{textcomp}
\usepackage[short]{optidef}
\usepackage{xcolor}
\usepackage{verbatim}
\usepackage{multirow}
\usepackage{svg}
\usepackage{bbding}
\usepackage{subcaption}

\usepackage{lipsum}

\def\BibTeX{{\rm B\kern-.05em{\sc i\kern-.025em b}\kern-.08em
    T\kern-.1667em\lower.7ex\hbox{E}\kern-.125emX}}
    
\begin{document}

\title{ZK-DPPS: A Zero-Knowledge Decentralised Data Sharing and Processing Middleware}

\author{Amir Jabbari\inst{1} \and
Gowri Ramachandran\inst{1} \and
Sidra Malik\inst{2} \and
Raja Jurdak\inst{1}}
\authorrunning{A. Jabbari, G. Ramachandran et al.}
\titlerunning{ZK-DPPS}
\institute{Trusted Networks Lab, Queensland University of Technology \\
\email\{{jabbaria; r.jurdak; g.ramachandran\}@qut.edu.au}
\and
Data61, CSIRO \\
\email{sidra.malik@data61.csiro.au}\\
}
\maketitle

\begin{abstract}
In the current digital landscape, supply chains have transformed into complex networks driven by the Internet of Things (IoT), necessitating enhanced data sharing and processing capabilities to ensure traceability and transparency. Leveraging Blockchain technology in IoT applications advances reliability and transparency in near-real-time insight extraction processes. However, it raises significant concerns regarding data privacy. Existing privacy-preserving approaches often rely on Smart Contracts for automation and Zero Knowledge Proofs (ZKP) for privacy. However, apart from being inflexible in adopting system changes while effectively protecting data confidentiality, these approaches introduce significant computational expenses and overheads that make them impractical for dynamic supply chain environments. To address these challenges, we propose ZK-DPPS, a framework that ensures zero-knowledge communications without the need for traditional ZKPs. In ZK-DPPS, privacy is preserved through a combination of Fully Homomorphic Encryption (FHE) for computations and Secure Multi-Party Computations (SMPC) for key reconstruction. To ensure that the raw data remains private throughout the entire process, we use FHE to execute computations directly on encrypted data. The "zero-knowledge" aspect of ZK-DPPS refers to the system's ability to process and share data insights without exposing sensitive information, thus offering a practical and efficient alternative to ZKP-based methods.
We demonstrate the efficacy of ZK-DPPS through a simulated supply chain scenario, showcasing its ability to tackle the dual challenges of privacy preservation and computational trust in decentralised environments.
\end{abstract}

\keywords{IoT Applications \and Blockchain \and Zero-Knowledge \and Publish-Subscribe \and Privacy-Preserving \and Smart Contracts \and Data Processing.}

\section{Introduction}
Integrating digital technologies has significantly enhanced data-driven decision-making and operational efficiency in the rapidly evolving supply chain management landscape. However, this transformation has escalated privacy risks, particularly for data producers and consumers who share sensitive information such as production capabilities and transaction details. The exposure of such data without adequate safeguards can lead to economic losses, competitive disadvantages, and regulatory non-compliance. Ensuring data confidentiality and integrity is thus crucial for maintaining trust and competitiveness in supply chains.
\par The use of Internet of Things (IoT) logistics transforms traditional supply chains into digital ones, providing real-time data on asset conditions and locations and linking physical assets with their virtual counterparts. The Publish-Subscribe (Pub-Sub) messaging paradigm is often employed to facilitate this link via secure communication between supply chain entities. The Pub-Sub model enables encrypted, loosely coupled communication between data producers and consumers, restricting unauthorised access within the system \cite{eugster2003many}. However, the broker-centred nature of traditional Pub-Sub introduces significant trust assumptions regarding data privacy. When data processing features are integrated into such centralised systems, as seen in solutions like Publish-Process-Subscribe (PPS \cite{centPPS}), privacy risks extend beyond data sharing to the processing itself. Reliance on centralised brokers for processing or on centralised computers for computation introduces vulnerabilities, as these components do not inherently preserve privacy.
\par In some implementations, the blockchain that is known for its decentralised and immutable nature has been integrated into Pub-Sub architectures to further enhance transparency and traceability \cite{ramachandran2019trinity}. However, there is an inherent tension between transparency and privacy. While transparency is essential for verifying the integrity of transactions and fostering trust among stakeholders, it should not compromise sensitive data privacy. Decentralised systems like Decentralised Publish-Process-Subscribe (DPPS \cite{dppsjabbari2024}) distribute computations across multiple nodes to mitigate single points of failure and leverage blockchain for enhanced transparency. Nevertheless, this decentralisation introduces additional layers where privacy risks can manifest. Therefore, there is a need for an approach that balances transparency with both data privacy and computation privacy, ensuring that distributed processing does not expose sensitive information.
\par Current privacy-preserving approaches often leverage blockchain and smart contracts, frequently relying on Zero-Knowledge Proofs (ZKP) for privacy-preserving data processing \cite{attema2020practical},\cite{sun2021survey}. However, these methods face significant limitations, such as smart contracts' inflexibility in handling computational changes and cost inefficiencies in fast-paced data-sharing environments \cite{zarir2021developing},\cite{ivanov2023security}. Integrating ZKPs within smart contracts introduces substantial computational overheads, including long setup and proof generation times \cite{solomon2023smartfhe}. Additionally, ZKP-based systems rely on trust in a single entity to generate and verify the proofs without tampering \cite{bowe2020zexe}. This centralisation of trust introduces a potential vulnerability, as a compromised proof generator could invalidate the entire process, undermining the privacy guarantees \cite{li2024ratel}. Furthermore, many of these solutions require a complete setup restart whenever computational changes are needed, reducing system flexibility and increasing costs \cite{steffen2022zeestar}. These constraints render existing state-of-the-art methods impractical for competitive supply chains demanding rapid data processing.
\par To address these challenges, we propose ZK-DPPS, a decentralised privacy-preserving data-sharing and processing middleware that ensures zero-knowledge communications without the need for traditional ZKPs. ZK-DPPS enables the extraction of insights from encrypted data while simultaneously verifying the correctness of the processes. The architecture integrates a consensus mechanism for Distributed Key Generation (DKG), where a public key is generated, and the corresponding private key is split into multiple shares. For key reconstruction and decryption, ZK-DPPS uses Secure Multi-Party Computation (SMPC) combined with Verifiable Secret Sharing (VSS). Fully Homomorphic Encryption (FHE) is used to perform computations on encrypted data, while a Byzantine Fault Tolerant (BFT) verification mechanism ensures computation integrity by verifying that the hash of the results is consistent across multiple replicated computations. The "zero-knowledge" aspect of ZK-DPPS refers to the system's ability to process and share data insights without exposing sensitive information, offering a practical and efficient alternative to ZKP-based methods.
\par In this proof-of-concept design, ZK-DPPS provides insights without relying on a central entity and without imposing computational burdens on the underlying pub-sub data producers and consumers. Moreover, ZK-DPPS avoids incurring gas fees during the insight extraction process, executing computations in a privacy-preserving manner. Key contributions include:
\begin{itemize}
    \item We design ZK-DPPS with a decentralised framework that ensures end-to-end privacy during data processing and verifies computation correctness using a BFT mechanism, providing secure insights for data consumers.
    \item We leverage SMPC and incorporate secret sharing in ZK-DPPS's key generation and key reconstruction steps to ensure no single entity can tamper with data integrity or the processes to access raw data. 
    \item We evaluate ZK-DPPS using a proof-of-concept through a simulated supply chain use case and compare it with the state-of-the-art to demonstrate its efficiency in a representative supply chain scenario. 
    \end{itemize}
    Our results show that ZK-DPPS achieves verifiable, decentralised data processing with competitive performance, particularly in scenarios with moderate message rates, while maintaining strong privacy guarantees. When compared to state-of-the-art approaches, ZK-DPPS offers improved privacy preservation without the high transaction costs and performance delays typically associated with smart contracts and Zero Knowledge Proof-based systems.
 
\par The rest of the paper is structured as follows: Section II reviews the literature and state-of-the-art approaches. Section III defines the system architecture of the proposed ZK-DPPS framework, while Section IV details its layered architecture. A thorough security analysis of ZK-DPPS is provided in Section V, and implementation details are discussed in Section VI. Section VII presents the evaluation results, followed by the discussion and conclusion in Section VIII.

 \section{State of the Art and Related Work}
In this section, we review the state-of-the-art literature related to privacy-preserving data sharing and processing, with a focus on supply chain management contexts.
\par Several works have investigated privacy-preserving data processing and computation within decentralised systems, particularly in the context of blockchain and SMPC \cite{li2024ratel},\cite{qu2022blockchain}. For instance, the Pub-Sub paradigm has been identified as a robust architecture for secure communication channels, aiming to prevent unauthorised access through encrypted, loosely coupled messaging models. However, traditional Pub-Sub models do not support data processing, necessitating systems like PPS \cite{centPPS}. Despite addressing some gaps, PPS still suffers from centralisation issues, where a single entity manages the data processing, leading to potential privacy vulnerabilities \cite{dppsjabbari2024}.
\par To mitigate these concerns, researchers have explored blockchain-based pub-sub architectures with integrated privacy-preserving computation techniques. For instance, some studies incorporate Homomorphic Encryption (HE) techniques and ZKP into smart contracts to enable privacy-preserving computations \cite{solomon2023smartfhe}. HE allows computations to be performed directly on encrypted data without requiring decryption \cite{gentry2009fully}. However, these approaches are often hindered by high computational costs and long proof generation times associated with ZKP systems \cite{bowe2020zexe}. In dynamic environments where there is a need for various types of insights, limitations and the inflexibility of computations' deployments pose significant adoption challenges.
\par The SmartFHE framework, for example, uses FHE in the blockchain setting to support privacy-preserving smart contracts \cite{solomon2023smartfhe}. It addresses the challenges lightweight users face by offloading computation to miners rather than requiring users to perform complex cryptographic operations themselves. However, SmartFHE encounters concurrency issues and struggles with verification times, particularly for large-scale implementations.
ZeeStar, another relevant work, combines HE with Non-Interactive Zero-Knowledge Proofs (NIZK) to implement privacy-preserving smart contracts on public blockchains \cite{steffen2022zeestar}. ZeeStar enables developers to specify intuitive privacy requirements without manually instantiating cryptographic primitives, which makes it a practical choice for Ethereum-based applications. Despite its practicality, ZeeStar's reliance on HE and NIZK results in significant gas fees and extended transaction times, making it less suitable for real-time processing in supply chains.
\par In light of these challenges, the ZK-DPPS framework proposed in this paper introduces a decentralised privacy-preserving data processing solution that leverages SMPC, FHE, and BFT-based verification to ensure data privacy and correctness without incurring the heavy computational costs seen in existing systems. By decentralising the data processing and verification processes, ZK-DPPS addresses the limitations found in previous approaches, offering a more scalable and efficient solution for privacy-preserving data sharing in supply chain contexts. 

\section{System Architecture}
This section discusses the system overview and introduces the key actors of ZK-DPPS system. The system model of our proposed decentralised Privacy-Preserving middleware is also discussed in this section. 
\subsection{System Overview \label{sysoverviewsubsec}}
\begin{figure}[h!]
\centering
\includegraphics[width=0.8\textwidth]{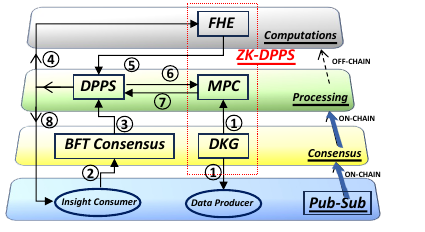}
\caption{The ZK-DPPS Architecture}
\label{TechInvolved}
\end{figure}

Figure~\ref{TechInvolved} shows the layered high-level architecture of ZK-DPPS and the integrated technologies in each layer for end-to-end privacy-preserving data processing. This architecture ensures that data consumers receive correct insights while maintaining data privacy through a carefully orchestrated series of steps.
The process begins in the consensus layer, where the system employs Distributed Key Generation (DKG) to collaboratively generate a public-private key pair. To mitigate the risks associated with centralised key generation, in the first step, the private key is split into multiple shares using VSS and distributed among relevant nodes, ensuring no single entity can reconstruct it. This design ensures that no single entity can reconstruct the private key, enhancing the system's security and decentralisation.
\par Once the keys are generated and distributed, the publisher encrypts the data using the generated public key. This encrypted data is then published to the network through the Publish-Subscribe paradigm, where it remains secure and accessible only in its encrypted form (step two). The communication between publishers, brokers (PPSManagers), and subscribers is secured through Transport Layer Security (TLS), ensuring that data is transmitted securely. ZK-DPPS achieves zero-knowledge communications by enabling the encrypted data to be processed and shared without revealing its content. This zero-knowledge Pub-Sub structure extends across all system communications, including those between publishers and PPSManagers, PPSManagers and distributed computers, and PPSManagers and subscribers. This consistent use of the zero-knowledge Pub-Sub structure ensures that sensitive information is protected throughout its entire life cycle within the system.
\par In the next phase, the processing layer facilitates decentralised data processing by leveraging DPPS system capabilities in step three \cite{dppsjabbari2024}. ZK-DPPS integrates privacy-preserving data processing features within the system to execute computations directly on ciphertexts using Fully Homomorphic Encryption (FHE). This allows for data processing without revealing the raw data, ensuring that sensitive information remains encrypted throughout the computational process (step four). Unlike systems that use Zero-Knowledge Proofs (ZKPs) to demonstrate the correctness of computations, ZK-DPPS achieves zero-knowledge communications by performing computations directly on encrypted data. This approach preserves data privacy by ensuring that only encrypted forms of data are handled during processing, thereby maintaining the confidentiality of the data at all times.
\par After the data processing phase, the computation layer performs cipher-to-cipher computations on the encrypted data. This ensures that the confidentiality of the data is preserved even during complex computational tasks. To verify the correctness of these computations, a Byzantine Fault Tolerant (BFT) verification process is employed (step five). The BFT verification involves validating the hash of the computation results across multiple nodes, ensuring that the computations are consistent and accurate. This verification process is essential for confirming the integrity of the results without compromising data privacy.\\ ZK-DPPS guarantees the privacy and correctness of computations as long as more than two-thirds of nodes in the consensus, processing, and computation layers are honest. This threshold ensures the system's resilience against Byzantine faults, making it robust even in adversarial settings.
\par Finally, the processed and verified encrypted results are prepared for decryption in the processing layer. ZK-DPPS uses Secure Multi-Party Computation (SMPC) protocols and Verifiable Secret Sharing (VSS) to securely reconstruct the private key necessary for decryption (step six). This key reconstruction is performed in a decentralised manner, ensuring that no single entity can access the private key in isolation, further reinforcing the system's zero-knowledge characteristics. By only involving the necessary number of nodes in this process, the system maintains privacy and security (Step seven). The decrypted insights are then delivered in a secure manner, ensuring that sensitive data is processed and shared without exposing raw data to potential risks (step eight).
\subsection{System Actors}
In ZK-DPPS, there are multiple entities interacting with each other, for which the key ones are introduced below.\\
\textbf{Publisher/Subscriber}: are the data producer and consumer in the network, publishing encrypted data to the network and receiving decrypted insights.\\
\textbf{PPSManagers (PPSM)}: act as the system's broker, facilitating communication and processing tasks between publishers, allocating tasks on the computers and interacting with the blockchain. A certain threshold of PPSManagers collaboratively construct the private key using the MPC protocols and VSS techniques.\\
\textbf{Validators (V)}: are the Consensus Nodes that validate transactions using a BFT protocol to ensure secure agreement on data integrity. Besides, they periodically generate sets of public key and shares of private key in a decentralised fashion.\\
\textbf{Computers}: Performs designated computations on encrypted data that are outsourced by PPSManagers, and generate traces along with the results for integrity consensus and later BFT-based correctness verification.\\
Each supply chain stakeholder participating in the network can nominate validators and/or PPSManagers in the system. Their incentive to contribute to the system is driven by the need to ensure the integrity and privacy of the network, while also benefiting from the ability to receive correct insights. Computers are also third-party computation provider such as cloud instances where they are hired at a monetary cost to outsource computational burden from traditional publishers and subscribers.

\subsection{System Interactions}
\begin{figure}[h!]
\centering
\includegraphics[width=1.0\textwidth]{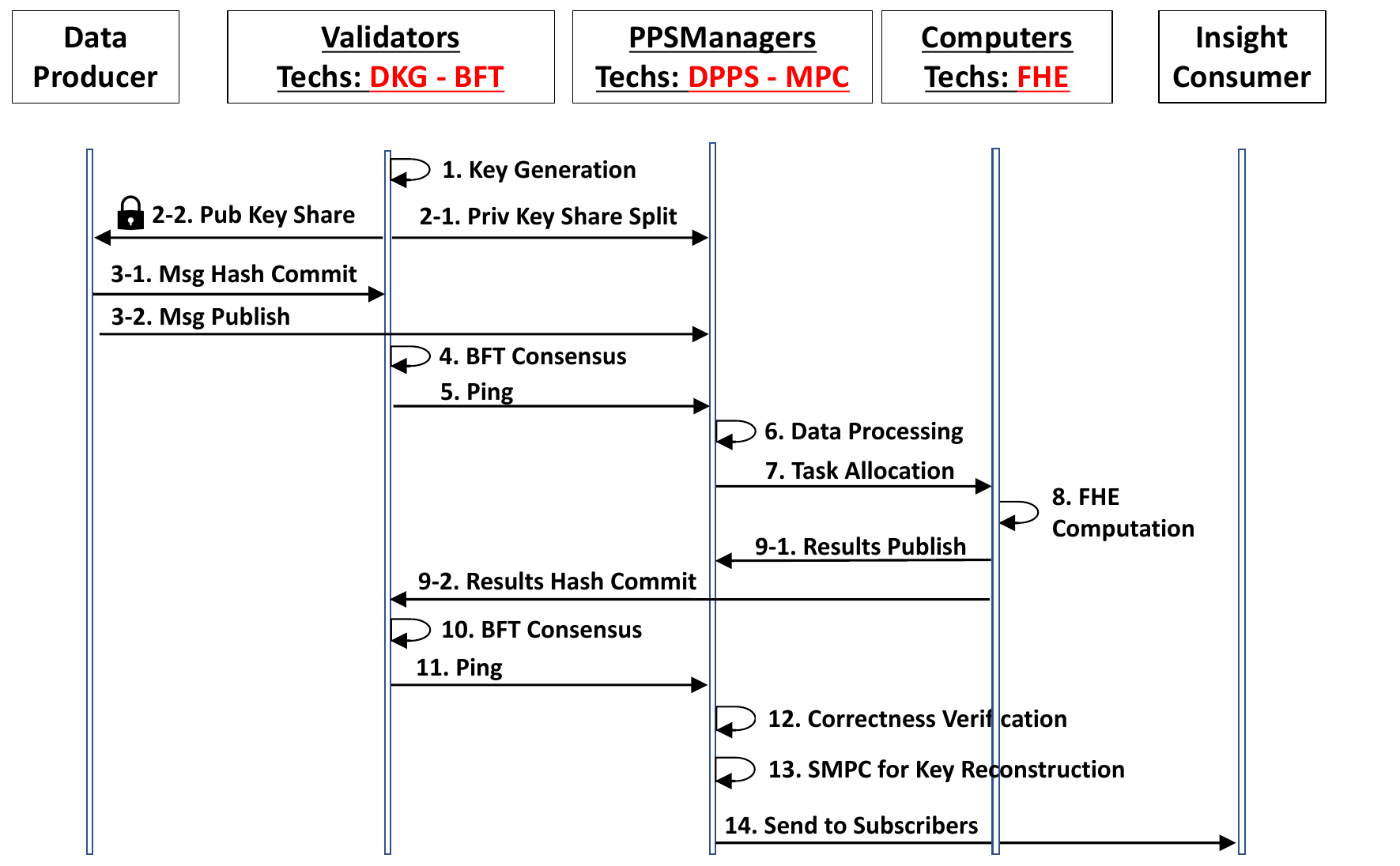}
\caption{Overview of Framework}
\label{seqsys}
\end{figure}
Figure~\ref{seqsys} presents an overview of the ZK-DPPS interactions. In this system, first, there is a round of Distributed Key Generation (DKG) within the Consensus Layer (step 1). Validators in the consensus layer periodically perform DKG to generate a new public key and shares of private key for each round of consensus. Private key is split into many shares, and the key shares are distributed among all PPSManagers to ensure no single PPSManager can reconstruct the key solely (step 2-2). The public key is distributed to publishers (step 2-2). Publishers encrypt their data using the public key generated by the DKG process and submit the hash of encrypted message to the consensus layer for commitment (step 3-1). The consensus layer commits the encrypted data to the blockchain, ensuring data integrity and consensus among validators. Publishers publish the message as the underlying pub-sub paradigm's requirement to their local PPSManager (step 3-2). After the consensus (step 4), all validators ping their interconnected PPSManagers for a new validated transaction (step 5). All the PPSManagers retrieve the validated transactions records, and match the hash of committed transaction with the hash included in the payload of received message to first confirm the integrity of data and second confirm the hash equality for the same published message (step 6).
PPSManagers then investigate the message topics \cite{centPPS} and once the intermediate processing steps (step 6) are facilitated by PPSManagers, PPSManagers aim for task allocations on random computers (\cite{dppsjabbari2024}) to execute computations on encrypted data (step 7). Distributed computers in the computation layer run the Cipher-to-Cipher functions on the encrypted data and results are committed to the consensus layer for being recorded on the ledger (step 9-2). Again, validators ping PPSManagers and PPSManagers run a BFT-based verification. For each computational result hash received in this verification method, there must be more than two-thirds of same hashes logged to the blockchain for same encrypted input (step 12). Once the computational results are verified, a random group of PPSManagers is selected to collaboratively reconstruct the private key for decryption purposes (step 13). This group is chosen using a secure random selection mechanism integrated within the consensus protocol, ensuring that the selection process is unpredictable and cannot be manipulated by any single entity. The selected PPSManagers then use their key shares to collaboratively reconstruct the private key in a privacy-preserving manner. After successful decryption, the results are shared with the subscriber and other stakeholders as needed (step 14).

\section{ZK-DPPS Layers}
As mentioned in Section~\ref{sysoverviewsubsec}, ZK-DPPS's architecture consists of multiple layers interacting together. 
The architecture of ZK-DPPS consists of two on-chain layers: consensus layer and decentralised processing layer, and one off-chain computation layer. These layers are discussed separately, followed by step-by-step description of the architecture.
\subsection{Consensus Layer \label{layer:cons}}
\begin{figure}[h!]
\centering
\includegraphics[width=0.8\textwidth]{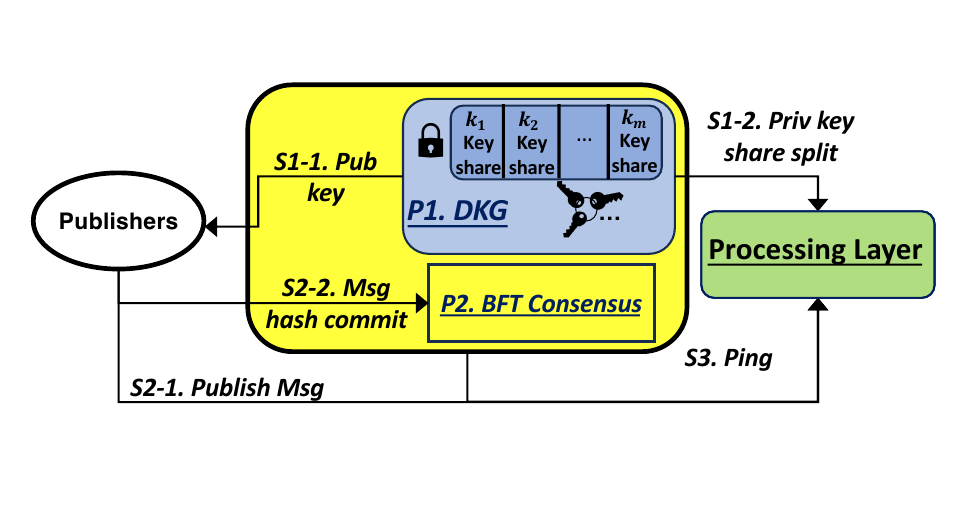}
\caption{Consensus Layer}
\label{layercons}
\end{figure}

In this layer, as demonstrated in Figure~\ref{layercons}, a BFT-based consensus happens on every incoming transaction where a majority of more than two-thirds of validators must come to a consensus for recording the transaction on the ledger. Validators also collaboratively participate in periodic key generation within the consensus layer. They generate the public/private key pair using a distributed key generation (DKG) protocol, such as FROST a Pedersen-based DKG, where each validator contributes to a shared public/private key pair generation \cite{komlo2021frost}. 
This protocol typically involves each validator generating a random polynomial and broadcasting commitments to the coefficients of this polynomial to all other validators \cite{pedersen1991threshold}. The DKG protocol consists of multiple phases, which we discuss below, to ensure secure and verifiable key generation processes among all the validators. \\
\textit{1) Initialisation Phase:} 
\par Each validator $ V_i $ generates a random polynomial $f_i(x)$ of degree $ t-1$, where $t$ is the threshold number of validators required to reconstruct the private key. The polynomial is of the form:
\begin{equation} \small
	\label{dkginit}
	f_i(x) = a_{i,0} + a_{i,1}x + a_{i,2}x^2 + \cdots + a_{i,t-1}x^{t-1}
\end{equation}
The constant term $a_{i,0}$ is the secret share of validator $ V_i $. \\
\textit{2) Commitment Phase} \\
Each validator commits to their polynomial $f_i(x)$ by broadcasting commitments to the coefficients $a_{i,j}$ to all other validators. The commitment for each coefficient $a_{i,j}$ is calculated using a cryptographic commitment scheme:
\begin{equation} \small
	\label{dkgcom}
	C_{i,j} = g^{a_{i,j}}
\end{equation}
where $g$ is a generator of the multiplicative group of a large prime order. \\
\textit{3) Distribution Phase} 
\par Each validator $ V_i $ securely sends the evaluation of their polynomial $f_i(x)$ at the points corresponding to other validators $ V_j $ (for $ j \neq i$ ). This evaluation $f_i(j)$ is kept private. Validators verify the received evaluations using the commitments. For $f_i(j)$, the verification is:
\begin{equation} \small
	\label{dkgdist}
	g^{f_i(j)} \stackrel{?}{=} \prod_{k=0}^{t-1} (C_{i,k})^{j^k}
\end{equation}

\par Periodically, to ensure the integrity of the DKG protocol and mitigate node collusion threat models, for validators that dispute an evaluation, they can request the disclosing validator to provide a zero-knowledge proof (ZKP) of the correctness of the evaluation. If the proof is invalid or not provided, the validator’s polynomial and its commitments are considered faulty, and they may be excluded from the process.
\par Once the protocol is executed, all the validators hold the same public key $P^{ub}$, with a share $i$ of $P^{riv}_{i}$ generated corresponding to the private key such that no set of shareholders smaller than the threshold knows the entire private key, $P^{riv}$. The private key $P^{riv}$ split in this protocol employs Feldman's Verifiable Secret Sharing (VSS) scheme to ensure that the shares distributions are correct and that the reconstruction process is verifiable \cite{komlo2021frost}. VSS helps to detect any malicious activity during the key generation and distribution phases \cite{feldman1987practical}.
\par Validators in the consensus layer are interconnected with a PPSManager from the processing layer. These pairs communicate securely using the mentioned established protocols. All communications between validators, PPSManagers, and registered computers are secured using the Transport Layer Security (TLS) protocol. TLS ensures data integrity, confidentiality, and authenticity by encrypting the messages exchanged \cite{eugster2003many}. Validators ping their corresponding PPSManagers with new key shares and new validated transactions using the secure communication channels between these two layers \cite{dppsjabbari2024}. 
\begin{figure}[h!]
\centering
\includegraphics[width=0.8\textwidth]{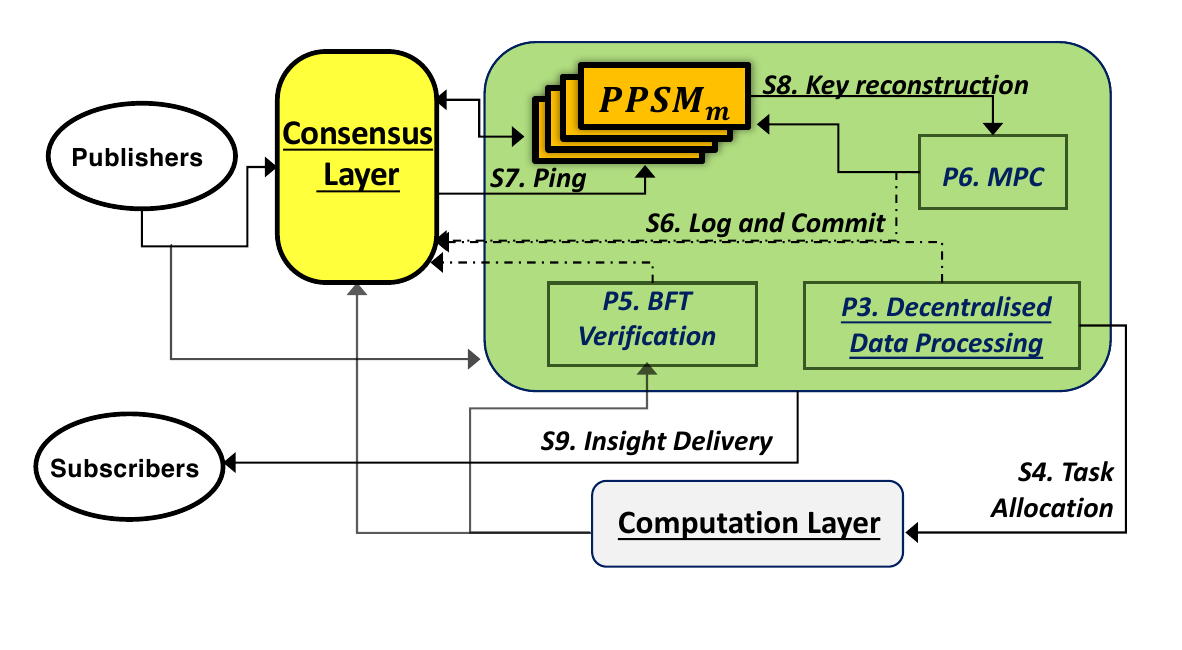}
\caption{Processing Layer}
\label{layerproc}
\end{figure}
\subsection{Decentralised Processing Layer \label{layer:process}}
PPSManagers in this layer are the connection facilitators between the consensus protocol and the computation layer. As shown in Figure~\ref{layerproc}, PPSManagers are connected to consensus nodes through the underlying blockchain platform's library and Application Blockchain Interface (ABCI), and are pinged whenever a new transaction is ready to be processed. After each ping that PPSManagers receive, they are triggered to confirm the existence of the records. This confirmation is performed by comparing the hash values included in the received message payloads with the key-value sets of the consensus transactions stored on the blockchain. Specifically, each message payload contains a serialised hash, which is a unique representation of the transaction's content, including the timestamp and other metadata. The serialised hash ensures that any alteration to the message will result in a different hash value, thereby safeguarding data integrity. PPSManagers retrieve the list of validated transactions from the consensus layer and match the hash of the committed transactions with the serialised hash included in the payload of the received message. By verifying that the hashes match, PPSManagers confirm both the integrity of the data and the consistency of the published message. This step is crucial for maintaining the correctness of the subsequent processing and ensuring that only valid transactions are processed further. Payload also includes the confidential raw data which is encrypted using the same public key $P^{ub}$ at the publishers' end.
\par To prepare the end-to-end privacy-preserving computation requirements for outsourcing, PPSManagers investigate the topic of received messages \cite{dppsjabbari2024}. This preparation includes multiple steps such as selecting a random computer and allocating tasks for pre-computation processes, and key partitioning and key reconstruction for the received computational results.
\\ \textit{1) Task Allocation}
\par Each PPSManager randomly selects a random computer from the pool of registered computers for cipher-to-cipher computation executions and allocate tasks for computations. This allocation follows the pub-Sub communication and the computer receives the encrypted raw data as a message payload along with the functions needed to run on the encrypted data. \\
\textit{2) Homomorphic Encryption Preparation}
\par To allow computations on encrypted data without decryption, homomorphic encryption schemes are used. For an encrypted message $ E(m)$, operations such as addition $ E(m_1 + m_2) $ can be performed:
\begin{equation} \small
	\label{homomorphs}
	E(m_1) \cdot E(m_2) = E(m_1 + m_2)
\end{equation}

\par Since the PPSManagers in Processing Layer are responsible for verifying the correctness of computations and decrypting results by reconstructing the private key using the Secure Multi-Party Computations (SMPC), several intermediate processes to ensure data privacy and computation integrity must be performed.\\
\textit{3) Key Share Partitioning}
\par To ensure key construction steps in a secure privacy-preserving manner where all the steps are audited to mitigate faulty PPSManagers collusion, the needed threshold of PPSManagers (e.g. $t$ of $m$) partition their key shares into smaller shares using techniques like Shamir’s Secret Sharing. This ensures that no single entity can access the complete key shares. The secret $S$ is split into $n$  shares $S_{i} $ such that any $t$ of them can reconstruct $S$ :
\begin{equation} \small
	\label{secretsharing}
	S(x) = a_0 + a_1 x + a_2 x^2 + \cdots + a_{t-1} x^{t-1}
\end{equation}
where each share $ S_{i} = S(x_{i}) $, with $x_{i}$ being a unique identifier for each participant. \\

\textit{5) Key Reconstruction} 
\par The private key $ P_{priv}$ can be reconstructed when needed by combining the shares from at least $t$ PPSManagers. The private key $ P_{priv}$ is reconstructed using Lagrange interpolation:
\begin{equation} \small
	\label{dkgdconstr}
	P_{priv} = \sum_{i \in T} \lambda_i f_i(0)
\end{equation}
where $T$ is the set of indices of the PPSManagers contributing their shares, and $ \lambda_i $ are the Lagrange coefficients:
\begin{equation} \small
	\label{dkgdconstr2}
	\lambda_i = \prod_{j \in T, j \neq i} \frac{j}{j-i}
\end{equation}

The correctness of the computation results is also verified where PPSManagers audit the hash of incoming computational results to validate the existence of more than two-thirds of same hash for each published data, ensuring the BFT property. 
These processes ensure the privacy and integrity of data throughout the computation life-cycle. By leveraging SMPC, the framework decentralises decryption and enhances security.

\subsection{Computation Layer: Secure Multi-Party Computations \label{layer:comp}}
\begin{figure}[h!]
\centering
\includegraphics[width=0.8\textwidth]{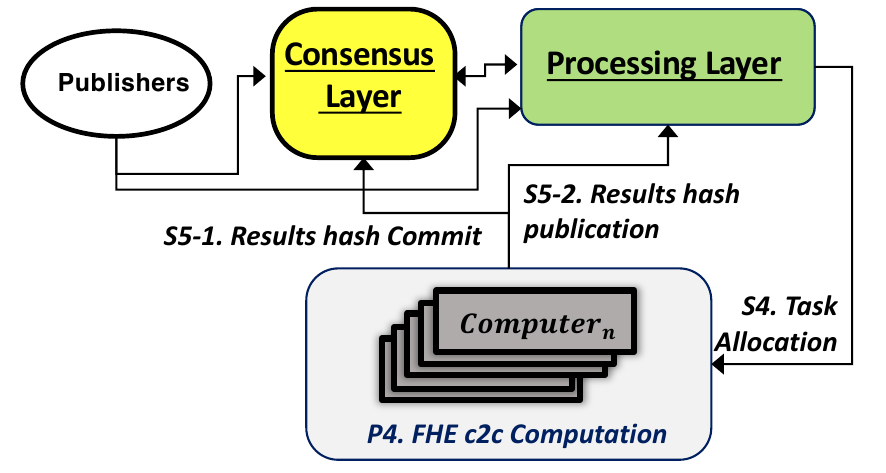}
\caption{Computation Layer}
\label{layercomp}
\end{figure}

A pool of registered, identifiable computers is in this layer. As mentioned before, the Computation Layer is responsible for executing tasks on encrypted data shares provided by the PPSManagers. As shown in Figure~\ref{layercomp}, this layer operates off-chain and each PPSManager in the network outsources tasks on randomly selected computers within the computation layer. Computations are executed to give encrypted shares without ever accessing the complete dataset, thus preserving data privacy at all times.
\par Computers perform computations on ciphertext using FHE techniques where there is no knowledge on the actual raw data. The concept of FHE is incorporated into the framework to significantly enhance privacy and security by distributing data processing across multiple computational nodes, thereby not only decentralising the handling of sensitive information but also enhancing the security of the framework. Key protocols are the same as discussion in section~\ref{layer:process} for the processing layer, and they include secret sharing of Equation~\ref{secretsharing} and homomorphic encryption of Equation~\ref{homomorphs}. After completing their computations, computers send the results to PPSManagers, and a random threshold of PPSManagers first reconstruct the private key using Equation~\ref{dkgdconstr2} and then aim to verify the correctness in a BFT fashion. 
\par The combination of SMPC and BFT-based verification of computations' correctness not only enhances the resilience of the competitive supply chains against data breaches and manipulation but also promotes a scalable and secure infrastructure for near-real-time data processing and decision-making.
\par In Figure~\ref{layercons}, once encrypted data is shared as a message by publishers as step one, the serialised hash of this message is committed as a transaction to the consensus layer in step two.
As discussed in Section~\ref{layer:process}, $PPSM_{m}$ and Validator $V{m}$ are interconnected using the ABCI. Therefore, once the block is created they are pinged for further processes. PPSMs facilitate HE processing requirements, and allocate encrypted data and related functions as tasks to computers and commit the logs as records for consensus by step four. Step five is another round of consensus for the committed logs. In step six, these allocated tasks by PPSMs on random computers are executed and the results of computations are returned for consensus in step seven. In step eight, the BFT-based consensus occurs on received computed share splits and using PPSMs key shares final aggregation and collaborative decryption is performed. In step nine the insights are delivered to subscribers with no privacy concerns for any publishers and the altering chance for any possible malicious parties participating in the network.

\section{Security Analysis and Threat Models}

This section delves into the security challenges and threat models confronting ZK-DPPS, outlining its defensive mechanisms against potential attacks. We use STRIDE categories to map the attacks to our proposed mitigation scenarios \cite{deng2011privacy}. We refer to the architecture Figures~\ref{layercons}, \ref{layerproc} and \ref{layercomp} to analyse threat models step by step. Along with handling the Man-in-the-middle (MITM) attacks, Resource exhaustion attacks, and Lazy Attacks by the underlying foundation of DPPS \cite{dppsjabbari2024}, the remaining scenarios are introduced below.  \\
\textbf{Multiple Computers Registration} is a threat when an attacker aims to register as many computers to the pool to take control of computations. In ZK-DPPS, privacy-preserving executions refer to the entire data processing pipeline, including insight extraction, where all computations are performed on end-to-end encrypted data. Using FHE, the raw data is never decrypted during processing. This ensures that even if an attacker attempts to register multiple computers in the network to gain control over the computations, they can never access the raw data. All computational steps, from aggregation to insight generation, are executed on ciphertexts, which prevents attackers from gaining any information even if they control multiple nodes. Moreover, the system leverages SMPC in key reconstructions, ensuring that no single node can control the entire private key reconstruction. This distributed approach further strengthens privacy, as malicious nodes cannot derive meaningful insights from the computations they perform on encrypted data.\\
\textbf{Sybil attacks}  involve an adversary creating numerous fake identities to overwhelm the network and manipulate consensus. In ZK-DPPS, Sybil attacks are mitigated by multiple layers of defence. Using secure communication channels, each communication channel is authenticated and encrypted. This ensures that only verified participants can join the network. The VSS protocol used in ZK-DPPS plays a critical role in mitigating Sybil attacks. Since only nodes with valid key shares can participate in key constructions, fake identities cannot meaningfully participate without the correct secret shares. Also, with encryption audit and validation, Each published message and computational task is associated with cryptographic proof of correctness. This validation step ensures that all participating nodes are properly authenticated, and their actions are verifiable.\\
These measures prevent attackers from creating multiple fake identities to influence the network or participate in computations. In addition, resource constraints are enforced through a credit-based publication system inherited from DPPS \cite{dppsjabbari2024}, limiting the ability of Sybil identities to flood the network with excessive data as \textbf{Distributed Denial of Service (DDoS)} Attacks. In ZK-DPPS, \textbf{DoS} attacks are mitigated by controlling the rate at which publishers can send messages to the network with a certain amount of credits allocated based on the principles of queuing theory \cite{dppsjabbari2024}. \\
\textbf{Node Collusion} occurs where multiple entities aim to collaboratively manipulate the network by sharing secret keys. ZK-DPPS mitigates this risk through periodic key generation using the DKG protocol. While it's true that an attack could theoretically occur before the keys are changed, several factors reduce this risk such as periodic key refreshing, threshold schemes and auditing with monitoring. The frequency of key changes is designed to be short enough that any successful collusion attack is minimised in time (the default in our evaluations is every 5 minutes). Once the keys are refreshed, previous collusion attempts become irrelevant. The threshold cryptography used in ZK-DPPS ensures that even if some nodes collude, they must control a significant portion of the network to succeed. Since key shares are distributed across many nodes, it is unlikely that a small subset of nodes can control the entire key without detection. he system also continuously audits and monitors the behavior of PPSManagers, ensuring that any suspicious activity (such as repeated failed computations or tampered messages) is flagged early. Overall, the dynamic nature of the DKG process means that even if nodes collude successfully at one point, they would have to consistently maintain control over the necessary threshold of key shares, which is difficult to achieve as the keys are periodically regenerated.\\
\textbf{Side-Channel Attacks} such as timing attacks, aim to exploit the time taken by operations to infer sensitive information about the cryptographic keys. In ZK-DPPS, the periodic DKG helps mitigate these attacks since the private key shares expire and are regenerated periodically.  Also, the use of SMPC ensures that no single node performs the full decryption process, which makes timing attacks more difficult to execute, as the critical operations are distributed across multiple nodes. Therefore, an attacker has a limited window of time in which to extract useful information, and even if the attacker gathers timing data from side-channel analysis, it becomes irrelevant after the next DKG round. \\
\textbf{Key Management Flaws} where there are possible flaws in the key generation, distribution, or management processes that could lead to key exposure or misuse, are mitigated using the employed VSS protocol mentioned in section~\ref{layer:cons}.

\begin{table*}[!htb]
\caption{ZK-DPPS Security Analysis Summary based on Processes in \textbf{Figure~\ref{seqsys}:}}
\label{tab:APIs}
\centering \tiny
\begin{tabular}{|c|c|c|c|c|}
\hline
\textbf{Attack} & \textbf{Targeted Layers} & \textbf{STRIDE Category} & \textbf{Risk} & \textbf{Mitigation}  \\
\hline
Key Management Flaws & Consensus Layer & Elevation of Privilege & Collude to reconstruct keys  & SMPC and VSS\\
Sybil Attack & Computation Layer & Information Disclosure & Malicious computer to reveal data & Homomorphic Encryption \\
Collusion & Processing Layer & Information Disclosure & Collude to decrypt raw data & Auditing and Logging \\
Lazy Attack & Computation Layer & Tampering & Modify data/computations & BFT-based Verification  \\
Consensus Protocol Attacks & Consensus Layer & Denial of Service & Disrupt the protocol and key generation & Robust Consensus Mechanism \\
DDoS & Processing Layer & Denial of Service & Message flooding to the system & Credit-based publication \cite{dppsjabbari2024} \\
Man in the Middle & Computation Layer & Spoofing & Unauthorised nodes claiming access & Encryption and Audits \\
\hline
\end{tabular}
\end{table*}

\section{Implementation Setup}
This section discusses the implementation and evaluation of  ZK-DPPS. The evaluation setup is introduced in Section~\ref{sec:evalsetup}. Section V-B discusses the use case applications. The performance evaluation is presented in Section V-C. 
\subsection{Evaluation Setup}\label{sec:evalsetup}
We implemented the ZK-DPPS using Mosquitto (MQTT) Broker, CometBFT\footnote{https://github.com/cometbft} Blockchain, MP-SPDZ \cite{mp-spdz} for DKG and VSS, and TenSeal \cite{tenseal2021} for Homomorphic Encryption and FHE computations. MQTT is one of the widely used publish-subscribe communication protocols in IoT applications. The broker-specific functionalities are implemented on top of MQTT, called PPSManagers in our middleware. For the blockchain, we used CometBFT, an open-source blockchain framework developed on top of the Tendermint Core. CometBFT blockchain framework consists of tools for achieving BFT-based consensus on a distributed network and the creation of blocks. The CometBFT consensus engine allows the application developers to replicate the state of an application across all the CometBFT validators in the network. The state information is fed into the consensus engine using the Application Blockchain Interface (ABCI). To implement the ZK-DPPS, we created a broker application using MQTT on top of the CometBFT blockchain framework. We used ABCI to bridge the MQTT application with the blockchain framework. The CometBFT framework uses Byzantine Fault Tolerance (BFT) consensus protocol as mentioned, which means 2/3 of the devices in the network must approve the transactions. When the majority of the devices in the network approve the transaction, the CometBFT framework adds the transaction in a block. All the application-specific software was implemented in NodeJS. We used Raspberry Pi 4 model B for evaluations, and we evaluated our setup on a Raspberry Pi test bed, including five Raspberry Pis. Each Pi is used as a separate domain and runs a validator and a PPSManager.
\subsection{Use Case Application}
In this experiment, we evaluate the system's functionality using the real-life supply chain use cases. In these scenarios, insights are needed to ensure the robustness of processes in a competitive supply chain, such as controlling the hazard threshold of carried goods, calculating the remaining life of the perishable products, and analysing the price fluctuations given the dynamic conditions. The correctness of insights in these scenarios can be critical since it might lead to unforeseen circumstances like human-life danger for spoiled food or product loss due to hazardous distribution. 
\subsection{Multi Publishers Scenarios}
Here are the mentioned scenarios for ZK-DPPS evaluations and experiments that align with our proposed framework's capabilities and the competitive supply chain context. For each of these scenarios, publishers, subscribers and the objective of each scenario is described as below.\\
\textbf{Scenario 1: Hazardous Material Transportation and Insurance}\\
Publishers are the RFID sensors of two companies (Publisher 1 and Publisher 2) that are transporting hazardous materials in the same truck.
Data that they publish are encrypted inputs about the quantities and types of hazardous materials, and they are published as messages.
Insurance Company is our third Publisher. The insurance company dynamically adjusts coverage based on the combined hazard levels of materials from both companies.
Subscribers in this scenario are also insurance regulators and the companies themselves receiving insights about the risk levels and insurance coverage.
The computations are cipher-to-cipher (c2c) summation and subtraction to calculate the total hazard level. \\
The objective is to demonstrate how ZK-DPPS processes encrypted data to ensure safe transportation and appropriate insurance coverage without revealing the exact nature of the materials.
\par \textbf{Scenario 2: Perishable Goods Shelf Life Prediction}\\
Publishers in this scenario are the temperature and humidity sensors of multiple suppliers that transport perishable goods such as fruits and vegetables.
Data that they publish are encrypted data on temperature and humidity conditions during transport.
Retailers and quality control departments are the subscribers who receive insights about the remaining shelf life of goods.
Computations executed in this scenario are c2c multiplications for calculating average conditions affecting shelf life.
The objective is to show how the ZK-DPPS framework can predict the remaining shelf life of perishable goods, ensuring quality control and reducing waste while maintaining data privacy.

\par \par \textbf{Scenario 3: Price Fluctuation Analysis in Supply Chains}\\
Multiple suppliers are the three publishers in this scenario that are dealing with a common product, such as electronic components.
They publish encrypted data on production costs, transportation costs, and selling prices.
Subscribers are market analysts and supply chain managers who want to receive insights about price changes and their causes.
There are c2c summation to aggregate costs, and
c2c multiplication for profit margin calculations.
The objective is to show how the ZK-DPPS framework can analyse price fluctuations, identify the sources of price increases, and provide actionable insights while preserving the confidentiality of each supplier's data.
Published data mentioned in each scenario is in form of messages and is published under a topic \cite{standard2014mqtt}. These topics determine the processes needed to undertake for each input value. Payload of each message must include the computation needed input parameter and the timestamp. Timestamp in ZK-DPPS works as a SALT for data processing and is a unique identifier for each message. Since the underlying messaging model is pub-sub based data sharing and processing, all the parties communicate through pub-sub and log and audit the hash of messages and actions to the blockchain. These hashes are also included in the message payload for transaction validation, and later the hash of computational results are used for BFT-based verification. Multiple scripts interact with each other on multiple machines. 
Here is a table regarding the message topics distributed in the system:
\begin{table}[!htb]
\caption{Message Topics}
\label{tab:messagetopics}
\centering 
\begin{tabular}{|c|c|c|}
\hline
\textbf{Role} & \textbf{Recipient Topic} & \textbf{Sent Topic} \\
\hline
Publisher & N/A & `` To-PPSM "  \\
\hline
PPSManager & `` To-PPSM " & `` To-Allocate "  \\
 &  `` To-Allocate " & `` To-Compute "  \\
 &  `` Results " & `` To-Decrypt "  \\
 &  `` To-Decrypt " & `` To-Subscriber "  \\
\hline
Computer & `` To-Compute " & `` Results " \\
\hline
Subscriber & `` To-Subscriber " & N/A \\
\hline
\end{tabular}
\end{table}

\section{Evaluation}
For the Distributed Key Generation (DKG) steps, as mentioned above, we used MP-SPDZ \cite{mp-spdz} with threshold requirements and Multi-Party Computation (MPC) techniques. MP-SPDZ is a framework for MPC that supports a variety of protocols and secret-sharing schemes and handles the necessary cryptographic operations for DKG with built-in support for Verifiable Secret Sharing (VSS). We leverage MP-SPDZ by employing Pedersen VSS along with Shamir's Secret Sharing in MP-SPDZ to ensure both security and verifiability of the secret-sharing process. Pedersen VSS can be used in conjunction with Shamir's Secret Sharing to ensure verifiable secret sharing, where each PPSManager can verify that the key shares they receive are consistent with the secret without revealing it.
\begin{algorithm}
  \caption{Key Generation and Distribution}\label{alg:keygen} 
  \begin{algorithmic}[1]
\STATE{\textbf{Input:} Number of PPSManagers $n$, Threshold $t$}
\STATE{\textbf{Output:} Distributed shares of the private key}
\STATE{Generate a random private key share $sk_i$}
\STATE{Create a secret share $S_i$ of $sk_i$}
\FOR{each PPSManager $P_j$ where $j \in [1, n]$}
    \STATE{Distribute the share $S_i$ to PPSManager $P_j$}
\ENDFOR
\STATE{Verify the validity of shares using Verifiable Secret Sharing (VSS)}
\IF{Shares are valid}
    \STATE{Output "Shares are valid."}
\ELSE
    \STATE{Output "Shares are invalid!"}
\ENDIF
\end{algorithmic}
\end{algorithm}
To generate the key set and distribute the private key shares, we implemented Algorithm~\ref{alg:keygen}, where we considered five total PPSManagers for distributing key shares and three PPSManagers as a threshold for reconstruction. ZK-DPPS's DKG codes are accessible in the ZK-DPPS\footnote{https://Link will be provided after acceptance} GitHub repository.
\par The execution time of Algorithm~\ref{alg:keygen} is 9 milliseconds on average. To reconstruct the private key using the key shares with the VSS method, we implemented Algorithm~\ref{keycon}, which took 21 milliseconds to execute.
\begin{algorithm}
  
  \caption{Key Reconstruction}\label{keycon} 
  \begin{algorithmic}[1]
\STATE{\textbf{Input:} Received shares $S_1, S_2, \dots, S_t$ from $t$ parties}
\STATE{ \textbf{Output:} Reconstructed private key $sk$}
\STATE{ Initialise $sk = 0$}
\FOR{each share $S_i$ where $i \in [1, t]$}
    \STATE{Compute Lagrange coefficient $\lambda_i$ from Equation~\ref{dkgdconstr2}}
    \STATE{ $sk \gets sk + S_i \times \lambda_i$}
\ENDFOR
\STATE{ Output the reconstructed key $sk$}
\end{algorithmic}
\end{algorithm}
\par The evaluation of ZK-DPPS shows its capability to achieve zero-knowledge communications with lower computational overheads compared to zero-knowledge proof (ZKP)-based systems. While ZKPs are effective for verifying computations without revealing data, they introduce significant processing delays and costs. In contrast, ZK-DPPS uses FHE for computations and employs MPC only for key reconstruction, avoiding the inefficiencies associated with ZKPs.
\par This design enables ZK-DPPS to securely process data with minimal overhead. As demonstrated by Algorithms~\ref{alg:keygen} and \ref{keycon}, the key generation and reconstruction are efficiently managed in a decentralised manner. Compared to solutions like ZeeStar, which incur high gas fees and execution times due to ZKP complexity, ZK-DPPS offers a more efficient and cost-effective alternative for privacy focused supply chain environments.

\subsection{End-to-End Delay}
\begin{figure}[hbt!]
\centering
\includegraphics[width=0.8\textwidth]{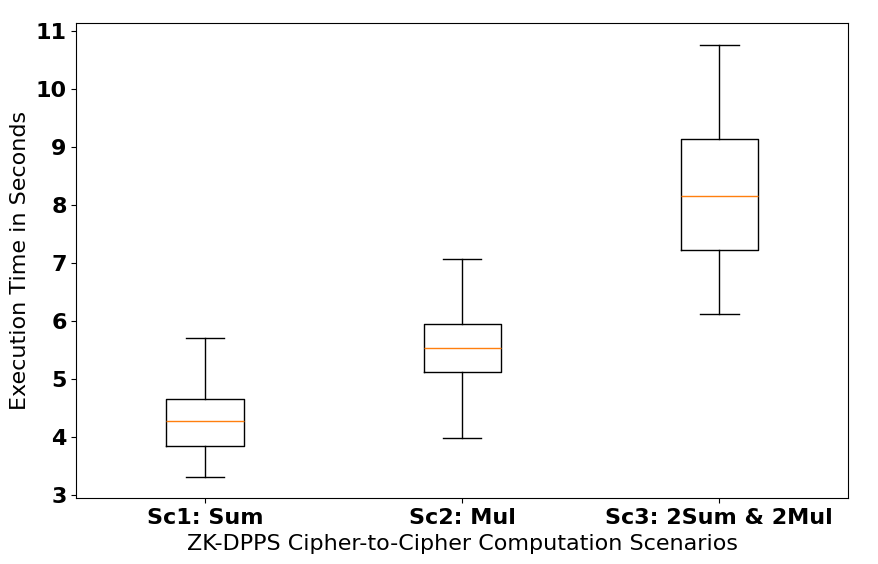}
\caption{End-to-End Delay of ZK-DPPS Evaluation Scenarios}
\label{baselines}
\end{figure}

The execution time for ZK-DPPS, which introduces cipher-text to cipher-text (c2c) computations in the evaluated scenarios, was carefully analysed. In these multi-publisher multi-operations scenarios, we aim to outsource computations to a random third-party computer to ensure the integrity of privacy-preserving computations and mitigate any collusion opportunity for learning publishers' encrypted data. As depicted in Figure~\ref{baselines}, our evaluations represent multi-publisher scenarios where each publisher publishes one encrypted message in the same mempool period. The blockchain's mempool period is configured for one block per second, which impacts the execution time of results since the transactions must be processed through validation by PPSManagers at each data processing step and final BFT verification. PPSManagers process published messages, and their encrypted inputs are set to be executed by a random computer that facilitates FHE computations.
Figure~\ref{baselines} demonstrates end-to-end delay of c2c tasks. These results are crucial for supply chain environments where timely processing of encrypted data is essential, showing that ZK-DPPS effectively maintains the balance between performance and privacy-preserving computations.

\subsection{Computation Cost Comparison}
To compare the evaluation costs against state-of-the-art privacy-preserving approaches, we use ZeeStar \cite{steffen2022zeestar} as a baseline. ZeeStar employs HE on Smart Contracts, using the Zkay programming language to automatically compile data privacy specifications into Ethereum smart contracts. This approach leverages HE and non-interactive zero-knowledge (NIZK) proofs. Executing a simple summation calculation with ZeeStar requires 339,000 gas units per transaction, which, as of 21/10/2024, equates to a minimum of 0.00678 ETH or approximately 18.37 USD per task computation. This demonstrates the significant computational cost associated with smart contract-based privacy-preserving data processing, which may become impractical in today's competitive supply chains.\\
\begin{figure}[h!]
\centering
\includegraphics[width=0.9\textwidth]{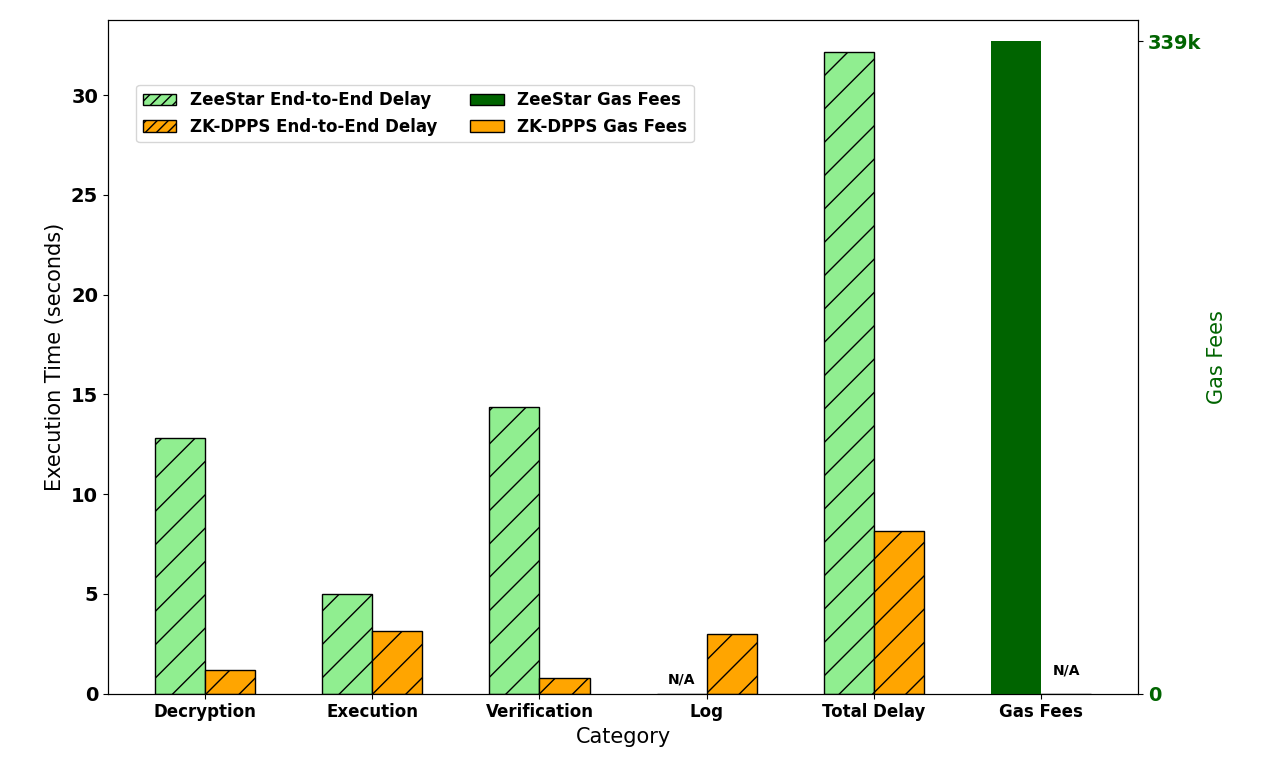}
\caption{\centering Scenario 3 Performance Comparison between ZK-DPPS and the state-of-the-art Privacy-Preserving Data Processing}
\label{d4psvszeestar}
\end{figure}
Additionally, ZeeStar’s approach takes approximately 54.7 seconds per transaction to generate NIZK proofs and perform HE computations. Depending on the complexity of the task and the encryption scheme, this duration could be even longer, overlooking ZeeStar's reported setup gas fees and overheads, which further add to the total computational cost. In contrast, ZK-DPPS avoids the burden of such gas fees while also providing significantly faster execution times.

Figure~\ref{d4psvszeestar} illustrates this comparison visually, showing that ZeeStar incurs both higher execution times (over 50 seconds) and gas fees (339,000 gas units) compared to ZK-DPPS, which requires neither gas fees nor extended execution times. The obvious differences in execution time and associated costs between ZeeStar and ZK-DPPS emphasises the efficiency and cost-effectiveness of ZK-DPPS, making it a more practical solution for data processing in supply chain scenarios.
\par For ZK-DPPS, the evaluations were conducted under normal operating conditions, with the system not under stress. The results represent the system's performance with an inter-message period of 5 seconds, allowing for a realistic assessment of its computation costs and efficiency. Due to the lack of direct access to ZeeStar's codebase for deployment and testing, we relied on the results published by Steffen et al. \cite{steffen2022zeestar}. Specifically, we referred to their reported end-to-end results for a scenario involving one cipher-to-cipher (c2c) summation and one c2c multiplication, including the time taken for decryption, deployment, and proof generation (Scenario 7).

\section{Performance Comparison}

\begin{figure}[h!]
  \centering
  \begin{subfigure}[b]{0.65\textwidth}
    \centering
    \includegraphics[width=\textwidth]{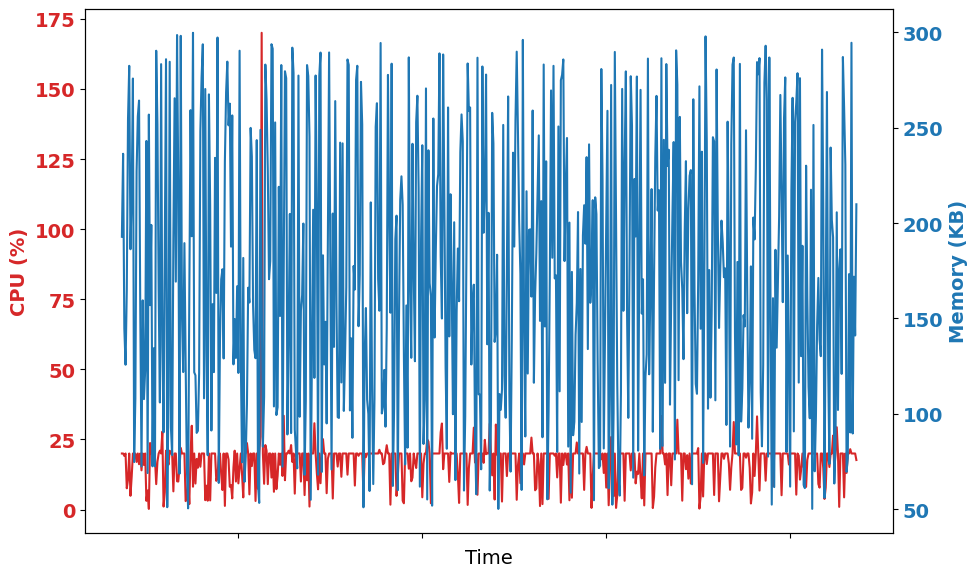}
    \caption{One Task per Half Second}
    \label{perfofig:fig1}
  \end{subfigure}
  \hfill
  \begin{subfigure}[b]{0.65\textwidth}
    \centering
    \includegraphics[width=\textwidth]{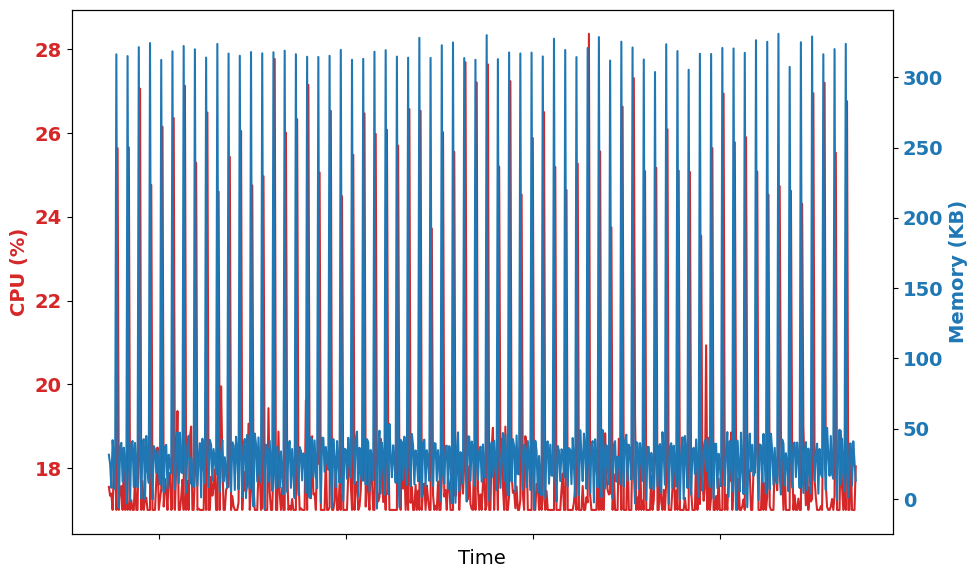}
    \caption{One Task per One Second}
    \label{perfofig:fig2}
  \end{subfigure}
  
  \vskip\baselineskip
  
  \begin{subfigure}[b]{0.65\textwidth}
    \centering
    \includegraphics[width=\textwidth]{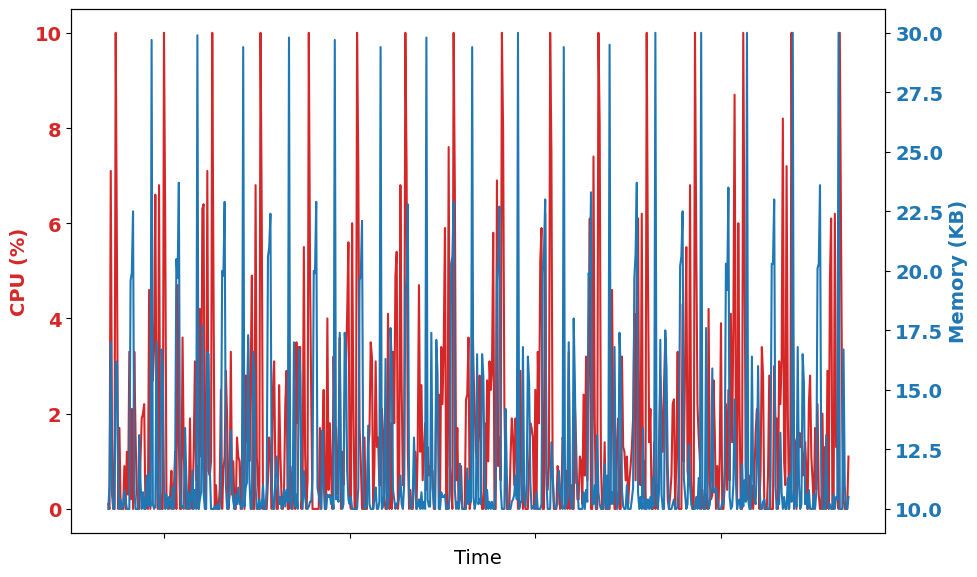}
    \caption{One Task per Five Seconds}
    \label{perfofig:fig3}
  \end{subfigure}
  \hfill
  
  \caption{CPU and Memory Performance of ZK-DPPS}
  \label{perfofig:main_figure}
\end{figure}

Figure~\ref{perfofig:main_figure} provides a detailed performance analysis of the PPSManagers’ (broker) resource usage under varying load conditions in ZK-DPPS, from the highest load in Figure~\ref{perfofig:fig1} to a lower load Figure~\ref{perfofig:fig3}. As the task interval increases, the system's memory and CPU consumption decreases, with ZK-DPPS exhibiting a clear advantage in resource efficiency compared to ZKP-based frameworks like ZeeStar \cite{steffen2022zeestar}. In ZKP-based solutions, the proof generation requires substantial computational resources, with memory requirements ranging from around 1 GB to 3 GB during the setup and compilation phase as evaluated in \cite{steffen2022zeestar}. This is due to the need to generate complex proof circuits and handle cryptographic operations like encryption, decryption, and homomorphic additions. These operations involve large arithmetic circuits, leading to significant memory consumption for rank-1 constraint systems (R1CS), as seen in ZeeStar's implementation \cite{9152634}. A rank-1 constraint system (R1CS) is a mathematical representation used in zero-knowledge proofs to express computational problems as a series of equations, enabling efficient verification of complex proof circuits by converting operations into linear constraints \cite{baumann2020zkay}.
\par While ZKP-based solutions, such as ZeeStar, require extensive memory and computational power for verification processes, ZK-DPPS optimises performance by leveraging BFT-based correctness verification without the computational overhead typically associated with ZKPs and smart contracts. These results suggest that ZK-DPPS is more suitable for resource-constrained environments, offering a privacy-preserving, decentralised architecture without the prohibitive resource costs of traditional ZKP methods. The decryption phase in ZK-DPPS also uses SMPC techniques and executes decentralised to provide correctness-guaranteed insights extracted by executing privacy-preserving computations.

\subsection{Stream Processing}
\begin{figure}[hbt!]
\centering
\includegraphics[width=0.8\textwidth]{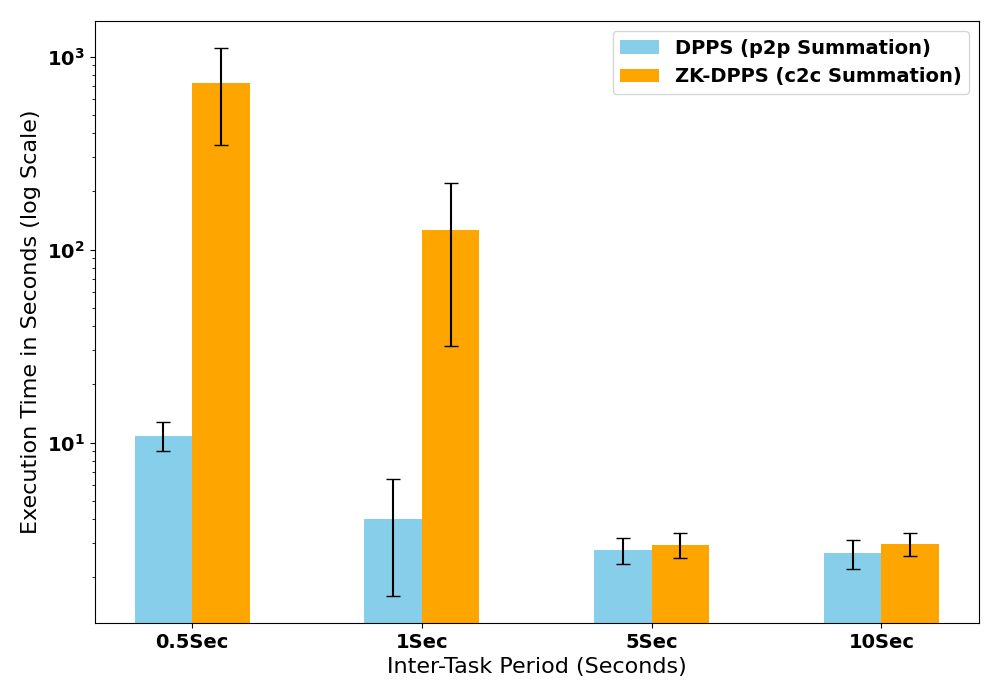}
\caption{\centering DPPS and ZK-DPPS End-to-End Execution Time Comparison for one Summation per Inter-Task Period (Seconds)}
\label{streaming}
\end{figure}

We compared the proposed ZK-DPPS with DPPS, a decentralised data sharing and processing middleware where privacy preservation is not a concern. DPPS is designed for stream processing in a supply chain context, enabling the computation of many tasks within short intervals. Figure~\ref{streaming} compares the end-to-end delay of DPPS and ZK-DPPS task execution times across different intervals.
\par While DPPS handles high-frequency tasks efficiently, ZK-DPPS incurs a noticeable spike in execution time as the number of published messages and subsequent tasks allocated by PPSManagers increases. This spike reflects the cost of incorporating privacy-preserving measures in a multi-stakeholder supply chain setting. However, for higher inter-task periods (i.e., when messages are published less frequently), the relative cost increase of ZK-DPPS over DPPS is significantly reduced. For instance, at a 10-second inter-task period, the added overhead of ZK-DPPS is on average 8\% higher than DPPS, demonstrating that the privacy-preserving mechanisms become less of a performance bottleneck in scenarios with lower message rates.\\
For ZK-DPPS, the evaluation involved varying the inter-task period in the range [0.5, 1, 5, 10] seconds over 10,000 runs for each interval. Although ZK-DPPS effectively guarantees end-to-end privacy-preserving data processing, its stream processing capability is challenged when faced with tasks that require execution at very short intervals, such as every second. The delay results from long message queuing and execution times, indicating that while ZK-DPPS scales well for privacy-focused applications, it may not be an ideal solution for use cases requiring ultra-low-latency stream processing. \\
This evaluation highlights ZK-DPPS's balance between scalability and privacy preservation. Unlike traditional stream processing systems, ZK-DPPS ensures that all computations are handled securely, introducing some performance trade-offs but remaining highly suitable for scenarios where privacy is paramount.

\section{Discussion and Conclusion}
ZK-DPPS demonstrates its effectiveness as a decentralised privacy-preserving middleware tailored for competitive supply chains, balancing security, scalability, and efficiency. Through its use of FHE for end-to-end encrypted computations and MPC for secure key reconstruction, it achieves a secure processing environment with low-latency key generation and reconstruction processes. Unlike state-of-the-art solutions like ZeeStar, ZK-DPPS eliminates the need for expensive smart contracts and the associated computational overheads, delivering a more cost-effective and faster solution for privacy-critical data processing.
\par However, ZK-DPPS is not without its limitations. While it performs well under conditions where the incoming message rate is balanced with the system's processing capacity, such as in scenarios with multi-publishers where encrypted data is published at intervals (e.g., every 1 seconds), it is not optimised for real-time or near-real-time processing. The increased end-to-end delay is a trade-off for maintaining privacy, making ZK-DPPS more suited for applications where privacy is paramount, but immediate insights are not required. In fast-paced environments that demand instant decision-making, the DPPS framework presented by the authors in \cite{dppsjabbari2024} may serve as a more suitable alternative.
\par Additionally, while ZK-DPPS integrates BFT and MPC to enhance security and ensure data privacy and integrity, these features introduce some computational overhead. ZK-DPPS relies on a BFT mechanism to ensure the integrity of computations across its consensus, processing, and computation layers. Provided that more than two-thirds of nodes in each layer are honest, ZK-DPPS can guarantee both the privacy and correctness of the results. Nevertheless, system performance may degrade under higher traffic or heavier workloads, highlighting the importance of identifying and maintaining an optimal threshold for message throughput to ensure real-world applicability.
\par Despite these trade-offs, ZK-DPPS provides a robust, secure, and scalable solution for decentralised data processing in multi-stakeholder environments. By prioritising privacy and offering a cost-effective alternative to ZKP-based systems, it addresses key challenges in supply chain data management. Future work will focus on optimising end-to-end delays and enhancing system efficiency to better support scenarios requiring both high privacy and faster processing. We will also investigate relaxing the BFT assumptions, experimenting with fewer honest nodes, to assess its impact on the system's performance and security.
\section{Acknowledgement}
This research is funded by iMOVE CRC and supported by the Cooperative Research Centres program, an Australian Government initiative [Project code: 5-033].

\bibliographystyle{splncs04}
\bibliography{zkdpps}

\end{document}